# Multiple sliding ferroelectricity of rhombohedral-stacked InSe for reconfigurable photovoltaics and imaging applications

Qingrong Liang[1], Guozhong Zheng[2], Liu Yang[3], Shoujun Zheng[1,]*

[1]Centre for Quantum Physics, Key Laboratory of Advanced Optoelectronic Quantum Architecture and Measurement (MOE), School of Physics, Beijing Institute of Technology, Beijing, 100081 China,
[2]School of Physics and Information Technology, Shaanxi Normal University, Xi'an, 710062, P. R. China
[3]Department of Physics, Hubei Engineering Research Center of Weak Magnetic-field Detection, China Three Gorges University, Yichang 443002, China

Corresponding authors: szheng@bit.edu.cn

**Abstract:** Through stacking engineering of two-dimensional (2D) materials, a switchable interface polarization can be generated through interlayer sliding, so called sliding ferroelectricity, which is advantageous over the traditional ferroelectricity due to ultra-thin thickness, high switching speed and low fatigue. However, 2D materials with intrinsic sliding ferroelectricity are still rare, with the exception of rhombohedral-stacked $MoS_2$, which limits sliding ferroelectricity for practical applications such as high-speed storage, photovoltaic, and neuromorphic computing. Here, we reported the observation of sliding ferroelectricity with multiple states in undoped rhombohedral-stacked InSe (γ-InSe) via dual-frequency resonance tracking piezoresponse force microscopy, scanning Kelvin probe microscopy and conductive atomic force microscopy. The tunable bulk photovoltaic effect via the electric field is achieved in the graphene/γ-InSe/graphene tunneling device with a photovoltaic current density of ~15 $mA/cm^2$, which is attributed to the multiple sliding steps in γ-InSe according to our theoretical calculations. The vdw tunneling device also features a high photo responsivity of ~255 A/W and a fast response time for real-time imaging. Our work not only enriches rhombohedral-stacked 2D materials for sliding ferroelectricity, but also sheds light on their potential for tunable photovoltaics and imaging applications.

**Keywords:** two-dimensional materials, van der Waals heterostructures, sliding ferroelectricity, γ-InSe, bulk photovoltaic effect, photodetection

# 1. Introduction

In the era of big data, exponential data growth requires innovative approaches to improve algorithms, storage and computational capabilities. Ferroelectric field effect transistors (FeFETs), which integrate the polarization switching of ferroelectrics into the nanodevice, offer significant potential to overcome the data processing bottlenecks in traditional von Neumann architecture[1-3]. Since the ferroelectric effect was first discovered in 1920, the family of ferroelectric materials has expanded significantly to include perovskite oxides, halide perovskites, and organic compounds, among others. The fertile materials have shown great potential for practical applications such as ferroelectric capacitors, FeFETs, and ferroelectric tunneling junctions[4-10]. However, traditional three-dimensional (3D) ferroelectric materials like $BiFeO_3$ have limitations due to size effects, strain effects, interface effects, and fatigue resistance, which constrain the miniaturization of FeFETs and large-scale applications[11].

Recently, two-dimensional (2D) ferroelectric materials are rapidly emerging[12, 13], particularly the sliding ferroelectrics, in which the spontaneous polarization resulting from the inversion symmetry breaking is caused by interlayer sliding and charge redistribution at van der Waals (vdW) interfaces. Compared to 3D ferroelectrics with switchable polarization by ion migration, the sliding ferroelectrics are characterized by large variety, small size and low polarization barrier. The sliding ferroelectricity was experimentally verified in vdW heterostructrues such as graphene/$h$-BN, $MoS_2$/$WS_2$, parallelly stacked $h$-BN and 1T'-$ReS_2$[14]. The unique interlayer sliding mechanism guarantees the high fatigue resistance demonstrated in the dual-gated 3R-$MoS_2$ device[15]. However, the mechanisms of interlayer sliding ferroelectricity in 2D materials are still unclear.

Although sliding ferroelectricity is theoretically predicted for some 2D materials with inversion symmetry breaking[16], it is difficult to experimentally observe the intrinsic sliding ferroelectricity in these 2D materials, with the exception of 3R-$MoS_2$. Theoretical work predicts that sliding ferroelectricity occurs in the rhombohedral-stacked InSe ($\gamma$-InSe)[3], which has a small band gap (1.0~1.6 eV) and high mobility (~$10^3$ $cm^2 \cdot V^{-1} \cdot S^{-1}$)[17]. However, sliding ferroelectricity is only reported in Yttrium-doped InSe[18, 19], where yttrium elements are used to modify the lattice constant and reduce intrinsic defects in $\gamma$-InSe. It is crucial to study the sliding ferroelectricity and elucidate the sliding ferroelectric mechanism in undoped γ-InSe, which can facilitate the integration of sliding ferroelectrics into practical optoelectronic applications.

In this study, we experimentally investigated the spontaneous polarization and multiple sliding switching states in undoped γ-InSe using dual-frequency resonance tracking piezoresponse force microscopy (Dart-PFM), scanning Kelvin probe microscopy (KPFM), and conductive atomic force microscopy (C-AFM). The multiple transition states guarantee the tunable bulk photovoltaic effect in graphene/ γ-InSe/graphene vdW heterostructures via an electric field[20]. Ab initio calculations elucidated the detailed sliding mechanisms with multiple routines in multilayer γ-InSe. The graphene/ γ-InSe/graphene device also features a high photo responsivity and fast response time for real-time imaging applications. Our work not only enriches the available sliding ferroelectrics to study the sliding mechanism, but also demonstrates the potential applications of 2D sliding ferroelectrics in photovoltaic and imaging devices, paving the way for the development of multi-bit optoelectronic devices.

## 2. Results and Discussion

### 2.1 Atomic structures and interlayer sliding of γ-InSe

γ-InSe has a hexagonal-rhombohedral crystal structure with lattice parameters a = b = 4.00 Å and c = 25.32 Å and belongsto the non-centrosymmetric space group R3m ($C^5_{3v}$). It has three armchair (AC) and three zigzag (ZZ) directions, as shown in **Fig. 1a-c.** Density functional theory (DFT) calculations predict sliding ferroelectricity in γ-InSe[21]. According to the Bulk Silicene Framework (BSF) theory, γ-InSe can exhibit both in-plane (IP) and out-of-plane (OOP) polarizations at certain high-symmetry points and along high-symmetry paths[22]. Taking the trilayer InSe as an example, three layers of γ-InSe are the ABC stacking state with a upside polarization (as shown in **Fig. 1a**). When the bottom layer is glided by 1/3 of the unit cell along the AC direction, it forms the antiferroelectric stacking state ABA with zero polarization (as shown in **Fig. 1b).** When the top layer is continuously glided by 1/3 of the unit cell along the AC direction, it forms the CBA stacking state with a downward polarization (as shown in **Fig. 1c**). DFT calculations show that three stacking states are energetically degenerate at room temperature (the energies of the metastable states CBA and ABA are only 0.002 meV and 0.003 meV higher than that of the ABC stacking state, respectively), suggesting the existence of multiple polarization states. Compared to the ABC structure, the layer spacings of the ABA and

CBA structures shrink by 0.12 and 0.14 pm, respectively (shown in **Figs.1a-c**). Therefore, the multiple polarization states in the multilayer γ-InSe can be achieved through interlayer sliding (Supplementary Fig. 1), which could be applied in multi-bit memory and optoelectronic devices.

We first performed the optical measurements of γ-InSe to confirm its atomic structures. Raman and photoluminescence (PL) spectra with different layer thicknesses confirm the R3m lattice structure and tunable bandgap properties (Supplementary Fig. 2-3 and Supplementary Note 1)[23, 24]. The layer-dependent second harmonic generation (SHG) intensity indicates the rhombohedral stacking nature of γ-InSe with the intrinsic inversion symmetry breaking (shown in **Fig. 1d**). The perfectly symmetric six-lobed pattern of SHG signal depends on the polarization angle of the excitation laser (**Fig. 1f**), suggesting the stacking order and the presence of OOP polarization. Furthermore, the absence of inversion and mirror symmetry allows charge transfer between adjacent atomic layers and spontaneous polarizations, which will be discussed in details later.

## 2.2 Multiple sliding ferroelectricity of γ-InSe

Sliding ferroelectricity in 2D materials is strongly influenced by stacking order, doping, interlayer coupling, etc., which requires advanced techniques to precisely study polarization switching. In previous studies, a high defectdensity was observed in the undoped γ-InSe[25] and the polarization value of γ-InSe is lower compared to other 2D sliding ferroelectric materials (e.g. h-BN, 3R-MoS$_2$, 1T'-ReS$_2$, 3R-LaBr$_2$ shown in Supplementary Table S1)[26] according to theoretical calculations. Furthermore, γ-InSe tends to be oxidized in air[27], which further limits ferroelectric characterization (Supplementary Note 2). Dart-PFM has proven to be a powerful tool for detecting ferroelectric polarization switching in ultrathin 2D nanoflakes[28]. The Dart-PFM mode enables the real-time detection of local ferroelectric polarization, greatly enhancing the sensitivity and accuracy of ferroelectric effects. To confirm the presence of ferroelectricity in γ-InSe, we first exfoliated flat γ-InSe flakes onto an Au-coated Si substrate (shown in **Fig. 2a** and Supplementary Fig. 4-5). Dart-PFM is then used to investigate the ferroelectric properties of γ-InSe using box-in-box domain engineering through an electric field. **Figure 2b** shows the OOP phase mapping of a 23 nm γ-InSe flake with box-in-box writing DC biases of +8 V and -8 V, clearly showing opposite polarization states. The sliding ferroelectricity of γ-

InSe is further confirmed by the PFM phase and amplitude loops in **Fig. 2c**. The amplitude loops exhibit a typical ferroelectric butterfly at room temperature with a pronounced 64° phase switching and an effective OOP piezoelectric coefficient $d_{33}$ of ~25 pm/V. We also tested other undoped γ-InSe flakes of different thicknesses and confirmed the presence of OOP ferroelectric polarization in undoped γ-InSe flakes[29] (see Supplementary Fig. 6-7).

To further verify the spontaneous polarization in γ-InSe, we measured the surface potential using the in situ KPFM mode[4]. The surface potential of a γ-InSe flake exhibits a uniform distribution in the absence of an external electric field, with the potential being about 300 meV lower than that of the Au substrate (as shown in **Fig. 2d** and Supplementary Fig. 8). A surface potential map is demonstrated in **Fig. 2e** when an electric field of +6 V (-6 V) was applied to the top (bottom) half of the sample. **Figure 2f** shows the surface potential plots along the white dashed line shown in **Figs. 2d-2e**, clearly demonstrating the potential difference between the top (~690 meV) and bottom (~-530 meV) halves of the sample. The clear surface potential contrast is still about 100 mV after removing the electric field in 30 min (see Supplementary Fig. 8), indicating the stability of the polarization switching in the undoped γ-InSe.

We also performed C-AFM measurements to examine the *I-V* curves of a 5 nm thick γ-InSe flake (see the atomic force microscopy (AFM) image in **Fig. 2g**). In the sweeping loop of the voltage from -4 to +4 V, *I-V* hysterisis can be clearly seen in **Fig. 2h**, which is attributed to the height modulation of Schottky barrier at the contact between the AFM tip and γ-InSe due to the polarization switching. Note that multiple current jump states appear as the voltage increases (see inset of **Fig. 2h**), indicating the presence of multiple polarization states in multilayer γ-InSe. According the DFT calculations, these jump states arise from multiple sliding ferroelectricity, and a schme of possible sliding states in tetralayer γ-InSe is shown in **Fig. 2i**. Furthermore, we found that distinct jumps were also observed for an 11.2 nm thick γ-InSe sample and the resistance states were significantly higher than for thiner samples (see Supplementary Fig. 9). These results demonstrate that sliding-induced multiple polarization states in multilayer γ-InSe can be modulated by applying an external bias voltage, revealing promising prospects for multi-bit storage and optoelectronics.

**2.3 Tunable bulk photovoltaic effect in γ-InSe**

The photovoltaic effect with reconfigurable optical response promises applications in low-power intelligent devices and image processing[30, 31]. The multiple ferroelectric polarization states of γ-InSe enable us to explore its tunable photovoltaic applications. We designed a graphene/γ-InSe/graphene sandwich structure (see **Fig. 3a**) to study bulk photovoltaic effect by monitoring the vertical current[32-34] (detailed fabrication process described in Method and Supplementary Fig. 10)[35, 36], where few-layer graphene flakes were used as bottom and top electrodes. The ultrathin γ-InSe with a thickness of 4.6 nm was confirmed by AFM and Raman spectroscopy (see **Fig. 3b**). Due to the high optical transmittance of graphene in the visible range[37, 38], our device enabled for investigating photogenerated carriers with simultaneous polarization of γ-InSe. We first performed short-circuit current ($I_{SC}$) mapping measurements on the vdW tunneling device under zero bias and found that the photogenerated current is only present in the overlap region of the top and bottom graphene electrodes (see **Fig. 3c**). Since the device was not prepolarized, the sign and magnitude of the photocurrent were unevenly distributed, which can be attributed to the spontaneous polarizations in γ-InSe. Current-voltage curves at different light power intensities are shown in **Fig. 3d**, indicating that the photovoltaic current is approximately linearly dependent on the light power output (see Supplementary Note 4 for the estimation of the power conversion efficiency of bulk photovoltaic devices).

We next investigated the tunability of the bulk photovoltaic effect in γ-InSe tunneling devices under electric field. The variation of the $I_{SC}$ under different poled voltage is summarized in **Fig. 3e.** When a +6 V (-6 V) poled voltage is applied, a negative (positive) $I_{SC}$ is present, indicating that the photovoltaic current is tunable according to the polarization states of γ-InSe. **Figure 3f** shows the switchable $I_{SC}$ of 15 nA and -19 nA at 22 μW light illumination. The reconfigurable photovoltaic current can be attributed to the tunable ferroelectric polarization via the electric field. The multiple polarization states due to the different stacking orders in the multilayer γ-InSe could be exploited to realize precise control of photovoltaic current via bias voltage or light illumination, which is promising for low-power intelligent applications.

### 2.4 Theoretical analysis of ferroelectric switching pathways in multilayer γ-InSe

To demonstrate the feasibility of the optimal sliding path for ferroelectric switching in γ-InSe, we referred to the generalized model proposed for 3R-MoS$_2$ from previous studies[15]. We performed DFT calculations to analyze the paths, structures, thermodynamics, kinetics, and polarization changes during the ferroelectric switching process in 2-, 3-, 4-, and 5-layer γ-InSe. Taking the trilayer γ-InSe as an example, ferroelectric switching can be achieved through three types of the interlayer sliding as shown in **Fig. 4a**. Regardless of the sliding type, atomic layers sliding along the AC direction is preferable for hexagonal lattice 2D materials[39-41]. In the first type (path 1), two adjacent atomic layers remain stationary while the last layer glides, resulting in a single sliding interface. In the second type (path 2), two non-adjacent atomic layers move in opposite directions while the middle layer remains stationary, resulting in two simultaneous sliding interfaces. In the third type (path 3), two adjacent atomic layers move in opposite directions while the last layer remains stationary, also resulting in two simultaneous sliding interfaces.

To identify the optimal sliding path among the three paths, we simulated the energy barriers for each path in γ-InSe. The energy profiles for the three paths differ for trilayer γ-InSe, with the path with minimum energy (shown in **Fig. 4b**). Intermediate polarization states (CBC and ABA) in path 1 have the energy barriers of 6 meV/f.u. in contrast to energy barriers of 12.02 meV/f.u. for path 2 and 86.3 meV/f.u. for path 3. The total sliding energy barrier is about 0.4 meV higher than that of the initial and final states, indicating the interlayer sliding along the path 1 most likely occurrs in multiple sliding ferroelectricity.

We extended this model to simulate the sliding steps for n-layer γ-InSe (n = 2 ~ 5) and found that n−1 energy barriers are formed in **Fig. 4c**. These findings allow us to analyze the switching process of an n-layer γ-InSe in n−1 steps. In each step, the n-layer γ-InSe is divided into a fixed block and a movable block by reversing the stacking sequence at the vdW interface. Because the interlayer repulsion at each sliding interface is independent, the subsequent sliding of each step is unrestricted and the polarization can switch to the several intermediate states. The intermediate polarization states exhibit different spontaneous polarizations depending on the number of previous switching steps as shown in Supplementary Fig. 14. These intermediate polarization states are thermodynamically stable and therefore can thus maintain their spontaneous polarizations (as shown in Supplementary Fig. 15). The number of intermediate polarization states for each spontaneous polarization in an n-layer system satisfies the condition k=n-1. Consequently, the total number of possible polarization states (N) in the entire ferroelectric switching process is $N=2^{n-1}$, supporting the multiple conductance states observed in C-AFM measurements. The exponential increase in the total number of polarization states with the number of layers, their thermodynamic stability, and different magnitudes of spontaneous polarization provide a blueprint for the development of sliding ferroelectric devices in the future.

To further confirm the reliability of the origin of vertical polarization, we simulated the Bader charge distribution and differential charge density distribution of γ-InSe polarization states for 2-5 layers (as shown in **Fig. 4d-I** and Supplementary Fig. 16). Bader charge analysis provides insights into symmetry breaking and interlayer charge transfer as illustrated in **Figs. 4d-e**. In CBA stacking, the net charges of the first, second, and third layers of InSe are -0.286, -0.204, and 0.491, respectively, while in ABC stacking, the net charges are reversed. **Figures 4f-g** show the different charge density and visually represent the interlayer electron cloud distribution of electron clouds between layers in CBA and ABC stackings. The electron cloud flow in the two different polarization states is clearly opposite, which is consistent with the experimental results of KPFM in **Fig. 2e**. **Figures 4h-i** show the planar averaged charge distributions ($\Delta\rho(z)$) along the c-axis for CBA and ABC stackings. To gain a deeper understanding of the relationship between sliding ferroelectricity and band structure changes, we simulate the differences in band structures between the ferroelectric and paraelectric states under spin-orbit coupling. This can also serve as a basis for switching the polarity state of sliding ferroelectricity (see Supplementary Fig. 17).

As the layer number increases, more stacking configurations become possible, potentially providing a sufficient number of polarization states for memristor switching devices. In contrast to previously reported bilayer ferroelectric systems, the parallel and antiparallel stacking configurations of γ-InSe multilayers are equivalent, allowing all stacking configurations to be exchanged by interlayer sliding. According to BSF theory, determining the presence of ferroelectricity in γ-InSe multilayers is straightforward: if both inversion and mirror symmetry are absent, the system is ferroelectric. Further details on the rapid determination of the presence of out-of-plane ferroelectricity in γ-InSe using BSF theory[22] can be found in Supplementary Note 5.

**2.5 Photo response performance of γ-InSe tunneling device**

To investigate the optoelectronic properties of the device, *I-V* curves of the graphene/γ-

InSe/graphene device were measured under various illumination intensities of 520 nm laser (as shown in **Fig. 5a** and Supplementary Fig. 18-S19) with a maximum photocurrent of 168 μA. The rise and fall times are obtained by an oscilloscope to be 388 μs and 538 μs, respectively (shown in **Fig. 5b**). This high photocurrent is attributed to the strong built-in electric field generated by the sliding ferroelectric γ-InSe between the graphene layers, improving the separation of photo-generated carriers (see **Fig. 5c** and Supplementary Note 6).

To further evaluate the optoelectronic performance of the graphene/γ-InSe/graphene vertical photodetector, we fabricated Au/γ-InSe/Au lateral photodetector to compare the photocurrent, responsivity (R) and photo-switching cycles of both devices. Responsivity was calculated using the equations R = $I_{ph}$/PS, where $I_{ph}$ is the net photocurrent, P is the power density, S is the effective illumination area[42-45]. **Figure 5d** shows the photocurrent of graphene/γ-InSe/graphene and Au/γ-InSe/Au photodetectors under the same bias. The vertical graphene/γ-InSe/graphene device exhibited a three order higher responsivity (255 A/W) than that of Au/γ-InSe/Au lateral device (127 mA/W), which can be attributed to the ferroelectricity induced built-in electric field and photogating effect. Photocurrent endurance measurements were carried out under periodically switching illuminations, as shown in **Fig. 5e**. Furthermore, the graphene/γ-InSe/graphene photodetector maintained a stable output photocurrent after 1200 cycles and showed excellent stability after 6 months (see Supplementary Fig. 20).

The high responsivity of vertical device is likely related to the interfacial effects between γ-InSe and graphene. The photocurrent is strongly affected by the built-in electric field between the top (bottom) graphene and γ-InSe layers. The built-in electric field arises from electron transfer from graphene to γ-InSe with the field direction from graphene to γ-InSe, as confirmed by our theoretical calculations (see Supplementary Fig. 21). Photogenerated electron-hole pairs can effectively separate and transport in the out-of-plane direction, which is absent in the Au/γ-InSe/Au lateral device. In addition, the long charge confinement lifetime of holes in γ-InSe leads to a significant photogating effect at the interface, where electrons circulate multiple times within the graphene, further improving the separation efficiency of photogenerated electrons and holes.

**Figure 5g** shows a schematic of the custom single-pixel imaging system, where a shadow mask with a hollow "BIT" block was placed behind a light source and moved in the x and y directions. When the photodetector received incident 520 nm light, the source meter recorded the photocurrent in real time and finally created a two-dimensional contrast map of the photocurrent intensity. The Gr/γ-InSe/Gr photodetector demonstrates the ability to accurately and clearly record and image the "BIT" logo, as shown in **Fig. 5h** and Supplementary Fig. 22, demonstrating the potential in real-time imaging.

## 3. Conclusion

In summary, we have reported the observation of sliding ferroelectricity in undoped rhombohedral-stacked γ-InSe using Dart-PFM technique and multiple ferroelectric states using C-AFM. The tunable photovoltaic effect is demonstrated in the graphene/γ-InSe/ graphene tunneling devices via the electric field and light illumination. According to theoretical calculations, the polarization switching of γ-InSe undergoes multiple sliding paths, which allows us to achieve a tunable bulk photovoltaic effect with a photovoltaic current density of 15 mA/cm$^2$. The vertical device also features a high photoresponsivity of 255 A/W and real-time imaging capability. Our work not only confirms the universal sliding ferroelectricity in rhombohedral-stacked 2D materials but also paves the path to explore sliding ferroelectricity for practical optoelectronic applications.

## 4.Methods
### γ-InSe Samples and vdW device preparation

γ-InSe nanosheets were mechanically exfoliated from bulk InSe using PDMS tape and adhered to an Au-evaporated Si substrate for PFM, KPFM, and C-AFM measurements. Raman mapping and photoluminescence (PL) of InSe nanosheets with different layer numbers are shown in Supplementary Fig. 1.

Bulk single-crystal graphite and bulk γ-phase InSe used in this study were purchased from 2D Semiconductors Inc. and Nanjing MKNANO Technologies. The number of layers in graphene and γ-InSe nanosheets are identified through optical images. After all materials were prepared, the vdW heterostructure devices were fabricated using an all-dry transfer technique. The flakes were sequentially picked up and placed onto glass slides using a PDMS stamp coated with a polyvinyl alcohol (PVA) film at 90 °C. The entire stacked structure was then released onto an SiO$_2$ (285 nm)/Si (p+-doped) substrate at 90 °C, followed by rinsing in deionized water.

Cr/Au (10 nm/50 nm) electrodes were fabricated on the stacked structures using electron beam lithography, followed by metal electron beam evaporation and lift-off processes, as detailed in Supplementary Fig. 9.

**Sample Characterization**

Optical imaging was performed using an Olympus microscope. Sample thickness measurements were conducted with an AFM in TappingMode in Air mode. Raman spectra and mapping were characterized using a WITEC alpha 200R Raman system with a 532 nm laser. SHG signals were measured using a horizontally polarized reflective geometry vertical microscope system equipped with a X100 objective (NA=0.5) and a spectrometer. A fiber pulse laser with a central wavelength of 1550 nm, a repetition rate of 18.5 MHz, and a pulse width of 8.8 ps was used as the fundamental pump radiation. Raman spectroscopy was conducted using a LabRAM HR Evolution spectrometer with a 532 nm laser. PL spectra were collected using 532 nm laser excitation.

**PFM, KPFM, and C-AFM Characterization**

PFM characterization of ferroelectric properties was performed using an Oxford atomic force microscope (Asylum MFD-3D Origin). PFM measurements were conducted in Dart-PFM mode with a Ti/Ir-coated conductive tip, a spring constant of 2.8 N/m, and a contact resonance frequency of approximately 320 kHz. Ferroelectric domain writing on InSe flakes was performed using ±9 V DC in LithoPFM mode of the Asylum Research software in contact resonance amplification mode, while reading the ferroelectric domains used a 1.5–2 V AC bias.

KPFM and C-AFM measurements were conducted using a Multimode 8 atomic force microscope (Bruker, USA). The sample probe had a Co/Cr conductive coating, a spring constant of 5 N/m, and a contact frequency of approximately 150 kHz. Compared to standard AC mode, KPFM included an additional feedback loop to record changes in surface potential. The first pass recorded surface information similarly to standard AC mode. In the second pass, the probe was lifted to a fixed height to record the potential difference between the probe tip and the sample.

**Photovoltaic Performance Testing of the Device**

Photocurrent mapping of the device was characterized using a Nanjing MStarter 200 (METATEST CORPORATION) photocurrent scanning test microscope. Electrical transport properties of the device were measured at a chamber pressure of $10^{-3}$ Torr at room temperature using an Agilent B1500A semiconductor device parameter analyzer.

**Theoretical Calculations**

First-principles calculations were performed using the Vienna Ab initio Simulation Package (VASP)[46-49], employing the Perdew-Burke-Ernzerhof generalized gradient approximation[50] for the exchange-correlation functional. Van der Waals corrections were included using Grimme's D3 method[51]. An energy cutoff of 550 eV was used for bilayer and trilayer models, and 600 eV for four-layer and five-layer models, with a Γ-centered k-point mesh of 15×15×1. The relaxation of atomic structures in each ferroelectric switching process was conducted with a force convergence criterion of 0.001 eV/Å on all atoms, and a convergence criterion of $10^{-7}$ eV for electronic self-consistent cycles. The climbing nudged elastic band method was used to calculate the geometries along the ferroelectric switching pathways[52], with a force convergence criterion of 0.01 eV/Å on all atoms. Calculations were performed with vertical vacuum separation of approximately 20 Å between periodic images of the 2D materials. The spontaneous polarization of the crystals was evaluated using the Berry phase method[53]. The spatial charge distribution of the systems was calculated using Bader charge analysis[54].

**Acknowledgements**

This work was supported by Natural Science Foundation of China grants (No. 12104050 and No. 62375018), National Key Research and Development Program of China (No. 2022YFA1203900) and Beijing Institute of Technology Research Fund Program for Young Scholars. The authors acknowledged Analysis & Testing center in Beijing Institute of Technology.

**Author contributions** Q.L. conducted the theoretical calculations, fabricated all the devices and obtained the data. G.Z assisted the theoretical calculations. Y.L. assisted experimental measurements. S.Z. supervised the project. All authors participated in scientific discussion and contributed on the manuscript.

lattice bias. *J Phys-Condens Mat,* **21**, 084204 (2009).

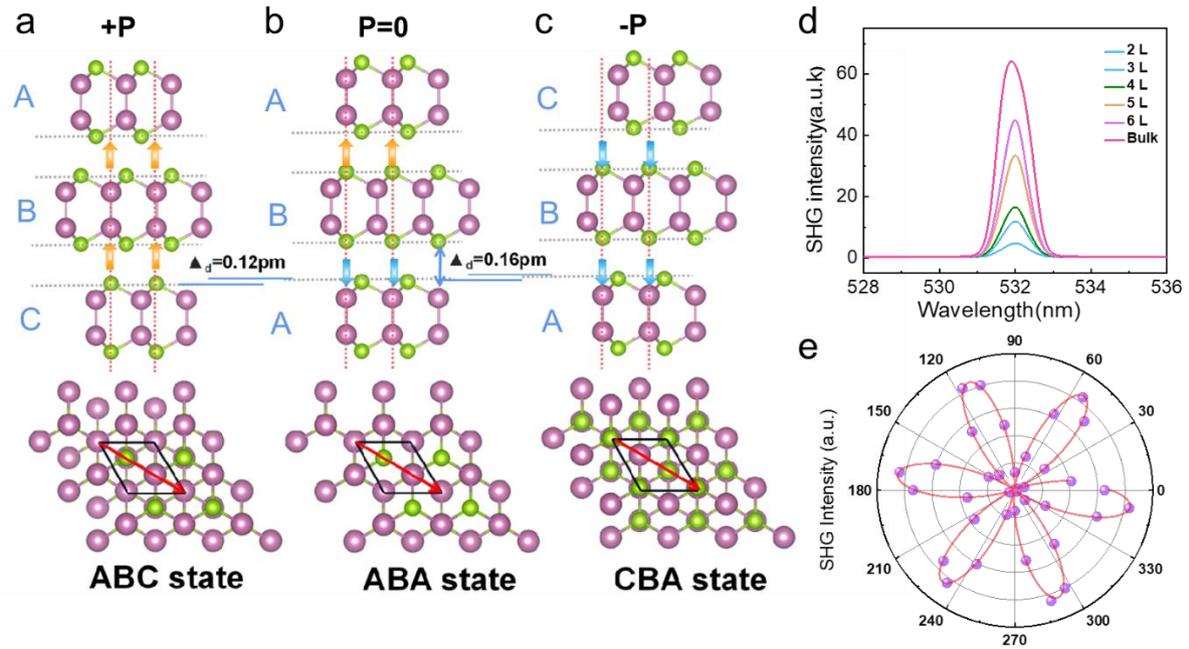

**Fig. 1 | Interlayer sliding of rhombohedral-stacked InSe. a-c.** Top views and side views of of ABC (**a**), ABA (**b**) and CBA (**c**) stacked γ-InSe. ABC stacking state can be switched to ABA and CBA stacking states via interlayer sliding, respectively. The opposite polarizations between ABC and CBA stacking orders indicates the existence of sliding ferroelectricity. **d**. Layer-dependent SHG spectra of γ-InSe indicating the rhombohedral stacking nature and inversion symmetry breaking of γ-InSe. **e**. SHG signal intensity as a function of the polarization angle with the fitted curve (red line).

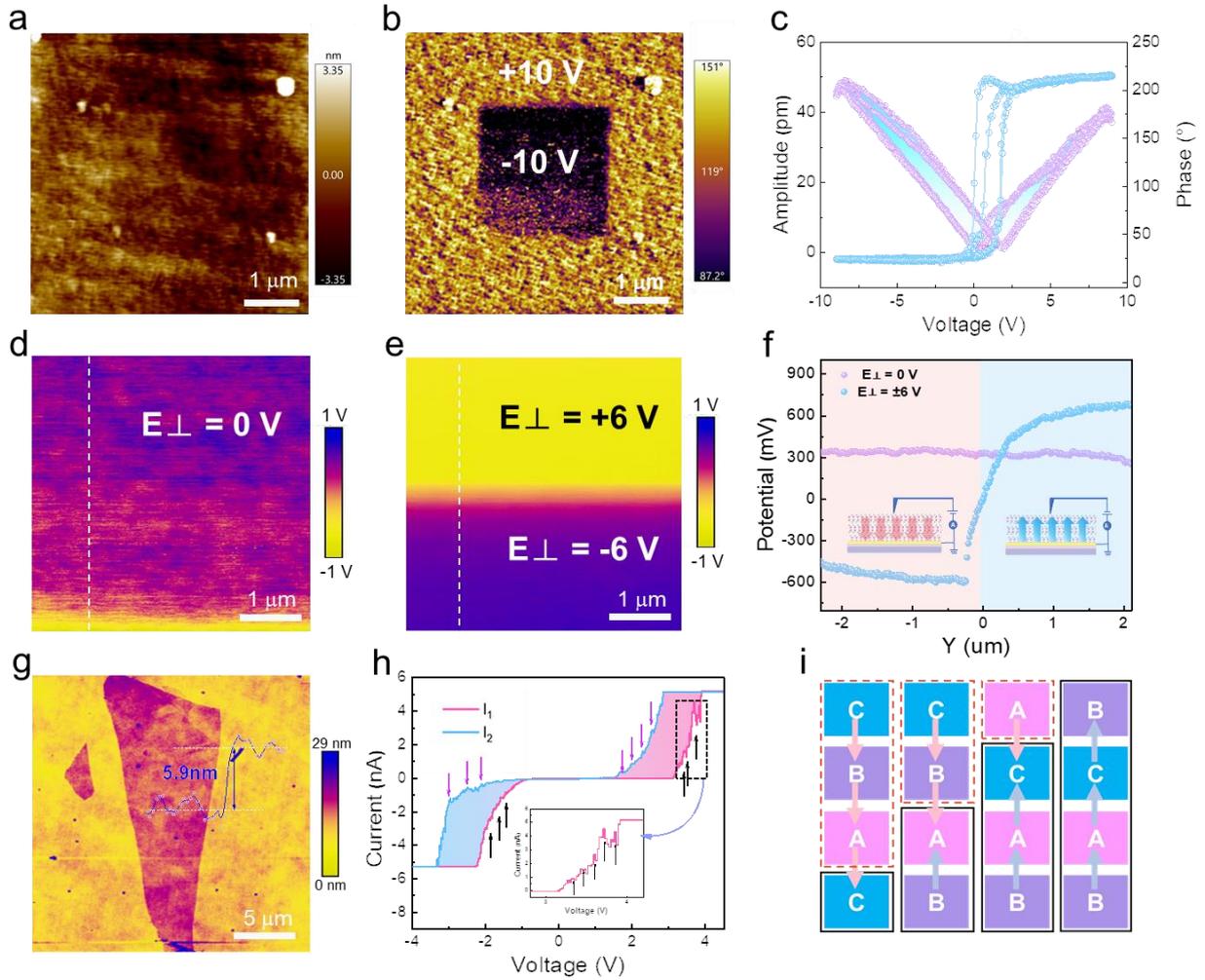

**Fig. 2 | Multiple sliding ferroelectricity of γ-InSe. a**. AFM morphology of a 23 nm thick γ-InSe flake. **b**. Phase mapping of the same region in (**a**) by a Dart-PFM with a box-in-box pattern (±10 V), clearly showing the switchable OOP ferroelectric phases. **c**. The PFM off-field phase and amplitude hysteresis loops with two cycles. **d**. KPFM mapping of a 10 nm thick γ-InSe flake before applying a vertical electric field. **e**. In-situ KPFM mapping of the same γ-InSe device in (**d**) under a ±6 V electric field. **f**. The plots of KPFM amplitudes along the white dashed lines showing the The and blue curves represent the surface potential variations in (**d**) (purple line) and (**e**) (blue line), respectively. **g**. Morphology of a 5.9 nm thick InSe flake used for C-AFM measurements (the thickness is shown in the inset). **h**. *I-V* curves of C-AFM domenstrating a clear hysterisis and multiple current jumps. The inset plot is zoom-in *I-V* curve demonstrating the existence of multiple polarization states. **I**. Schemetic of the multiple polarization states formation in a tetralayer γ-InSe via interlayer sliding.

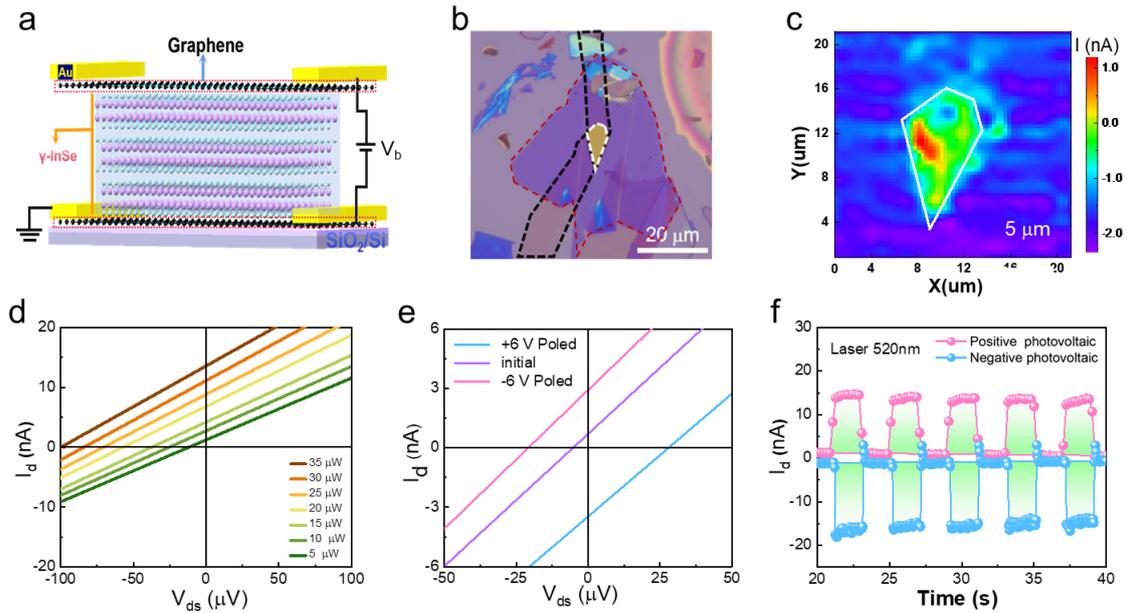

**Fig. 3 | Tunable Bulk Photovoltaic Effect in graphene/γ-InSe/graphene tunneling device.**
**a**. Schematic of the graphene/γ-InSe/graphene junction. **b**. Optical image of the tunneling device where the red dashed region is the γ-InSe flake, black dashed regions are the two graphene flakes and the white dashed region is the core region of the vdW junction for photovoltaic test. **c**. The $I_{SC}$ mapping of the tunneling device, clearly demonstrating the $I_{SC}$ originated from the core region of the tunneling device. **d**. I-V curves under various powers of 520 nm light illumination from 5 μW to 35 μW. **e**. Switchable photovoltaic behavior under different poling voltages, showing the electric field tunable $I_{SC}$. **f**. Endurance test of $I_{SC}$ with positive and negative $I_{SC}$.

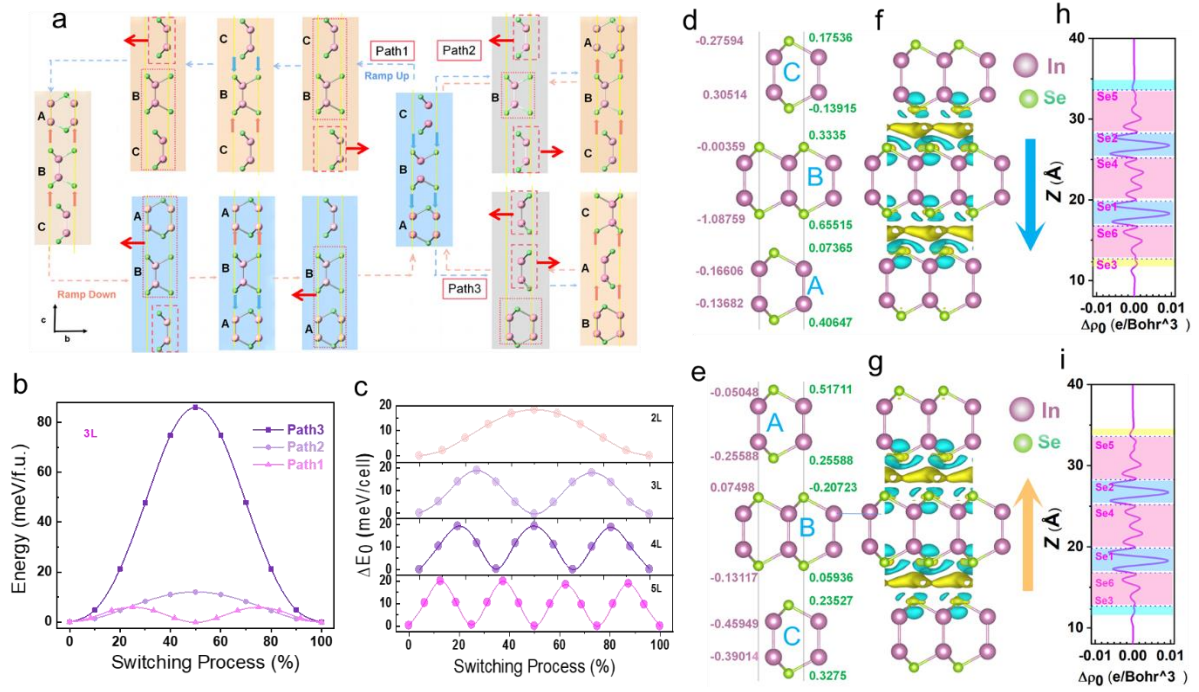

**Fig. 4 | Multiple ferroelectric switching pathways in γ-InSe. a.** Path 1, 2 and 3 of the polarization switching in a trilayer γ-InSe. Atomic structures at 0%, 25%, 50%, 75%, and 100% of the switching processes are shown. Red dashed (dotted) boxes indicate the moving (stationary) blocks. Gray horizontal lines indicate the sliding interfaces. The solid red horizontal arrows mark the displacement direction of the individual blocks. Atomic structures without solid red horizontal arrows are stable structures. Purple and blue vertical arrows indicate the spontaneous polarization directions of the interface dipoles. **b.** Total energy profiles of ferroelectric switching pathways of trilayer γ-InSe showing the energy profiles of paths 1, 2, and 3. **c.** Energy profiles of paths for bilayer, trilayer, tetralayer and pentalayer obtained from the generalized model. **d-e.** The Bader charge analysis of the CBA (**d**) and ABC (**e**) polarization states. **f-g.** The differential charge density distribution of CBA (**f**) and ABC (**g**) stacking states These images reveal the inequivalent distribution of the electron cloud after interlayer stacking with yellow and blue isosurfaces representing electron accumulation and depletion, respectively. **h-i.** Planar-averaged screening charge profiles among CBA (**h**) and ABC (**i**) stacking states.

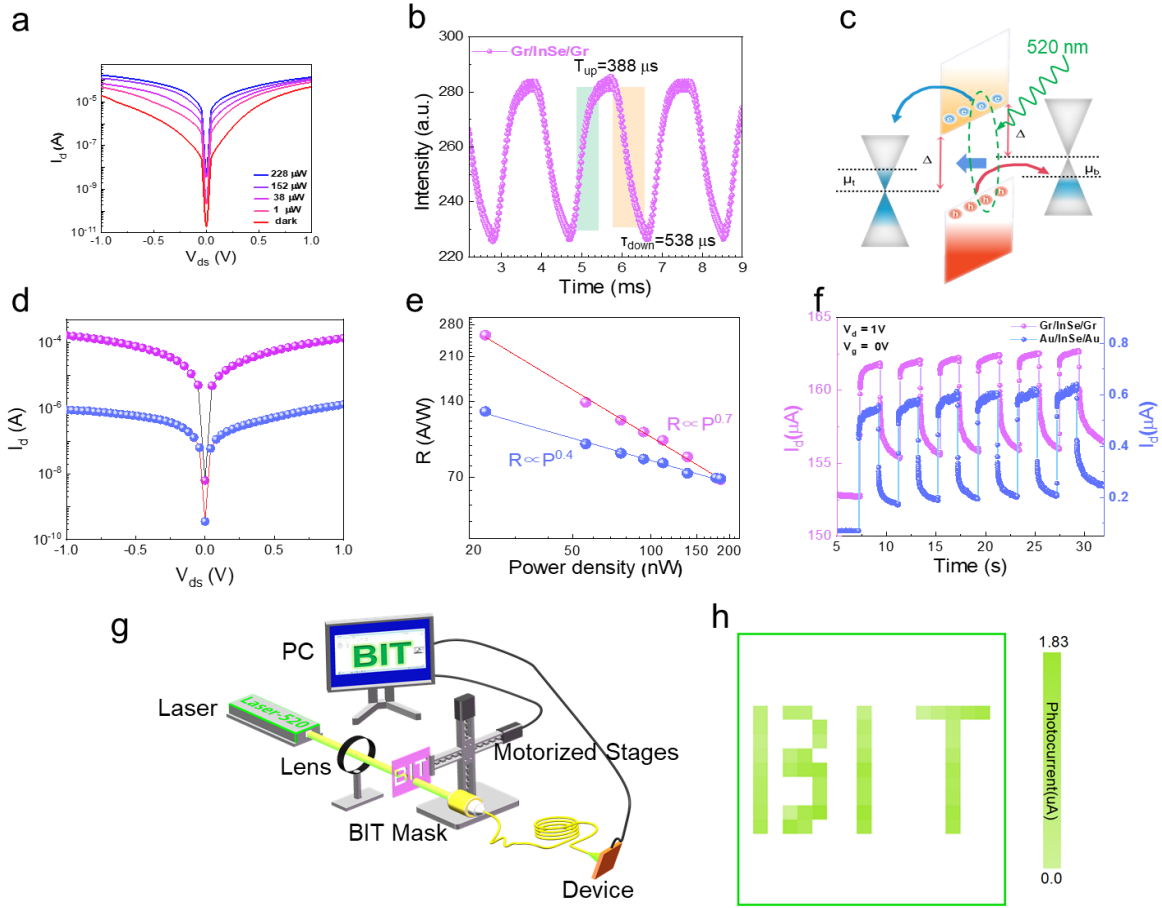

**Fig. 5 | Photo response performance of γ-InSe tunneling device a**. *I-V* curves of the graphene/γ-InSe/graphene vdW photodetector under various light intensities. **b** Photo response of the photodetector showing the rise time of 388 μs and decay time of 538 μs. **c**. Schematic of band alignment of the graphene/γ-InSe/graphene structure. **d**. Comparison of *I-V* curves between the graphene/γ-InSe/graphene vertical device (purple curve) and the Au/γ-InSe/Au lateral device (blue curve) under 25 μW light illumination. **e**. Responsivity as a function of power of two devices. **f**. Photocurrent endurance test of two devices at the applied voltage of 1 V. **g**. Schematic of real-time imaging based on graphene/γ-InSe/graphene photodetector. **h**. The "BIT" logo acquired from the single-pixel imaging system based on the 520 nm light irradiation.